\documentclass[12pt,fleqn]{article}
\usepackage{english,latexsym,amssymb,graphics}
\usepackage{epsfig,feynmp}

\makeatletter
\newif\if@preliminary
\@preliminaryfalse
\def\preliminary{\@preliminarytrue}
\def\preprintno#1{\def\@preprintno{#1}}
\def\address#1{\def\@address{\textit{#1}}}

\def\abstract#1{\def\@abstract{#1}}
\newlength\preprintnoskip
\newlength\abstractwidth
\@titlepagetrue
\renewcommand\maketitle{\begin{titlepage}%
  \let\footnotesize\small
  \hfill\parbox{\preprintnoskip}{%
  \begin{flushright}\@preprintno\end{flushright}}\hspace*{1cm}
  \vskip 60\p@
  \begin{center}%
    {\Large\bf\boldmath \@title \par}\vskip 1cm%
    {\sc\@author \par}\vskip 3mm%
    {\@address \par}%
    \if@preliminary
      \vskip 2cm {\large\sf PRELIMINARY DRAFT \par \@date}%
    \fi
  \end{center}\par
  \@thanks
  \vfill
  \begin{center}%
    \parbox{\abstractwidth}{\centerline{\bf\abstractname}%
    \vskip 3mm%
    \@abstract}
  \end{center}
  \end{titlepage}%
  \setcounter{footnote}{0}%
  \let\thanks\relax\let\maketitle\relax
  \gdef\@thanks{}\gdef\@author{}\gdef\@address{}%
  \gdef\@title{}\gdef\@abstract{}\gdef\@preprintno{}
}%
\def\@citex[#1]#2{\if@filesw\immediate\write\@auxout{\string\citation{#2}}\fi
  \def\@citea{}\@cite{\@for\@citeb:=#2\do
    {\@citea\def\@citea{,\penalty\@m}\@ifundefined
       {b@\@citeb}{{\bf ?}\@warning
       {Citation `\@citeb' on page \thepage \space undefined}}%
\hbox{\csname b@\@citeb\endcsname}}}{#1}}
\def\citerange{\@ifnextchar [{\@tempswatrue\@citexr}{\@tempswafalse\@citexr[]}}
\def\@citexr[#1]#2{\if@filesw\immediate\write\@auxout{\string\citation{#2}}\fi
  \def\@citea{}\@cite{\@for\@citeb:=#2\do
    {\@citea\def\@citea{--\penalty\@m}\@ifundefined
       {b@\@citeb}{{\bf ?}\@warning
       {Citation `\@citeb' on page \thepage \space undefined}}%
\hbox{\csname b@\@citeb\endcsname}}}{#1}}
\long\def\@makecaption#1#2{%
  \vskip\abovecaptionskip
  \sbox\@tempboxa{#1: \emph{#2}}%
  \ifdim \wd\@tempboxa >\hsize
    #1: \emph{#2}\par
  \else
    \hbox to\hsize{\hfil\box\@tempboxa\hfil}%
  \fi
  \vskip\belowcaptionskip}
\newcommand{\beq}{\begin{eqnarray}}
\newcommand{\eeq}{\end{eqnarray}}
\newcommand{\nn}{\noindent}
\newcommand{\non}{\nonumber}

\newcommand{\pskip}{\vspace{\baselineskip}}
\newcommand{\s}{\\ \vspace*{-3.5mm} } 
\newcommand{\qq}{$q\bar{q}$}

\makeatletter
\def\fmslash{\@ifnextchar[{\fmsl@sh}{\fmsl@sh[0mu]}}
\def\fmsl@sh[#1]#2{%
  \mathchoice
    {\@fmsl@sh\displaystyle{#1}{#2}}%
    {\@fmsl@sh\textstyle{#1}{#2}}%
    {\@fmsl@sh\scriptstyle{#1}{#2}}%
    {\@fmsl@sh\scriptscriptstyle{#1}{#2}}}
\def\@fmsl@sh#1#2#3{\m@th\ooalign{$\hfil#1\mkern#2/\hfil$\crcr$#1#3$}}
\makeatother
\def\fmfL(#1,#2,#3)#4{\put(#1,#2){\makebox(0,0)[#3]{#4}}}

\begin{document}
\baselineskip16pt   

\preprintno{%
hep-ph/9904287\\
DESY 99/033\\
TTP99-17\\
PM/99-21}

\title{%
 Production of Neutral Higgs-Boson Pairs at LHC
}
\author{%
 A.~Djouadi$^1$, W.~Kilian$^2$, M.~Muhlleitner$^3$ and 
 P.M.~Zerwas$^3$ 
}
\address{%
  $^1$Lab. de Physique Math\'{e}matique, 
        Universit\'{e} Montpellier, F-34095 Montpellier Cedex 5\\
  $^2$Institut f\"ur Theoretische Teilchenphysik, 
        Universit\"at Karlsruhe, D-76128 Karlsruhe\\
  $^3$Deutsches Elektronen-Synchrotron DESY, D-22603 Hamburg
}
\abstract{%
  The reconstruction of the Higgs potential in the Standard Model or
  supersymmetric theories demands the measurement of the trilinear
  Higgs couplings. These couplings affect the multiple production of
  Higgs bosons at high energy colliders. We present a systematic
  overview of the cross sections for the production of pairs of
  (light) neutral Higgs bosons at the LHC. The analysis is carried out
  for the Standard Model and its minimal supersymmetric extension.}
\maketitle
\subsection*{1. Introduction}

{\bf 1.}  Self-interactions of the Higgs field in the scalar sector
induce the breaking of the electroweak symmetry $SU(2)_L\times U(1)_Y$
down to the electromagnetic symmetry $U(1)_{EM}$ of the Standard Model
(SM). Gauge bosons and fermions acquire masses by interactions with
the non-zero Higgs field $v=1/(\sqrt{2} G_F)^{1/2}$ in the ground
state of the scalar potential. It is therefore an important
experimental task to reconstruct the elements of the Higgs potential
which gives rise to the spontaneous breaking of the electroweak
symmetry. The shape of the potential is determined by the mass $M_H$
of the physical Higgs boson field, and its trilinear and quadrilinear
couplings. The trilinear coupling \cite{hhhlc},
\beq
\lambda_{HHH} = 3M_H^2/M_Z^2
\eeq
in units of $\lambda_0 = M_Z^2/v$, can be measured directly in the
production of Higgs-boson pairs at high energy colliders.  In proton
collisions at the LHC, Higgs pairs can be produced through double
Higgs-strahlung off $W$ and $Z$ bosons \cite{hrad}, $WW$ and $ZZ$
fusion \cite{fusion}, and gluon-gluon fusion \cite{glover}; in generic
notation:
\beq
\begin{array}{l l l c l c l}
\mbox{double Higgs-strahlung}& \hspace{-0.3cm} : & q\bar{q} & 
\hspace{-0.3cm} \to &\hspace{-0.1cm}  W^*/Z^* &
\hspace{-0.3cm} \to &\hspace{-0.1cm}  W/Z + HH 
\non \\[0.1cm]
WW/ZZ\ \mbox{double-Higgs fusion}& \hspace{-0.3cm} : & qq & 
\hspace{-0.3cm} \to & \hspace{-0.1cm}  qq + WW/ZZ &
\hspace{-0.3cm} \to & \hspace{-0.1cm}  HH 
\non \\[0.1cm]
\mbox{gluon} \; \mbox{fusion} &  \hspace{-0.3cm} : & gg &
\hspace{-0.3cm} \to & \hspace{-0.1cm}  HH & & \non
\end{array} 
\eeq
Characteristic diagrams of the three processes are shown in
Fig.\ref{smdiag}. With values typically near 10~fb, high integrated 
luminosities are needed to generate a sufficiently large ensemble of 
signal events and to cope with the large number of background events. 
\pskip

\nn {\bf 2.}  The Minimal Supersymmetric extension of the Standard
Model (MSSM) incorporates a quintet of Higgs bosons: $h$, $H$, $A$,
$H^{\pm}$; the particles $h$, $H$ are neutral and CP-even while $A$ is
neutral and CP-odd. The mass of the light CP-even Higgs boson $h$ is
limited to less than about 130~GeV. The masses of the other Higgs
bosons are typically of the order of the electroweak symmetry breaking
scale $v$, yet they may extend up to values of order 1~TeV. The MSSM
Higgs system is described inherently by two parameters which are
generally chosen as the mass $M_A$ of the pseudoscalar Higgs boson and
the mixing parameter $\tan\beta$, ratio of the vacuum expectation
values of the two neutral Higgs fields. Radiative corrections
introduce the mass and mixing parameters of the heavy $t/ \tilde{t}$
and $b/\tilde{b}$ chiral multiplets into the system. \s

In CP-invariant theories, six types of trilinear couplings are
realized among the neutral Higgs fields \citerange{hhhlc,djouadi}:
\beq
\begin{array}{l l l l}
hhh, & Hhh, & HHh, & HHH \non \\
hAA, & HAA &     & \non
\end{array} 
\eeq
\begin{fmffile}{fd}
\begin{figure}[ht]
\begin{flushleft}
\underline{double Higgs-strahlung: $q\bar q\to ZHH/WHH$}\\[1.5\baselineskip]
{\footnotesize
\unitlength1mm
\hspace{10mm}
\begin{fmfshrink}{0.7}
\begin{fmfgraph*}(24,12)
  \fmfstraight
  \fmfleftn{i}{3} \fmfrightn{o}{3}
  \fmf{fermion}{i1,v1,i3}
  \fmflabel{$q$}{i1} \fmflabel{$\bar q$}{i3}
  \fmf{boson,lab=$W/Z$,lab.s=left,tens=3/2}{v1,v2}
  \fmf{boson}{v2,o3} \fmflabel{$W/Z$}{o3}
  \fmf{phantom}{v2,o1}
  \fmffreeze
  \fmf{dashes,lab=$H$,lab.s=right}{v2,v3} \fmf{dashes}{v3,o1}
  \fmffreeze
  \fmf{dashes}{v3,o2} 
  \fmflabel{$H$}{o2} \fmflabel{$H$}{o1}
  \fmfdot{v3}
\end{fmfgraph*}
\hspace{15mm}
\begin{fmfgraph*}(24,12)
  \fmfstraight
  \fmfleftn{i}{3} \fmfrightn{o}{3}
  \fmf{fermion}{i1,v1,i3}
  \fmf{boson,tens=3/2}{v1,v2}
  \fmf{dashes}{v2,o1} \fmflabel{$H$}{o1}
  \fmf{phantom}{v2,o3}
  \fmffreeze
  \fmf{boson}{v2,v3,o3} \fmflabel{$W/Z$}{o3}
  \fmffreeze
  \fmf{dashes}{v3,o2} 
  \fmflabel{$H$}{o2} \fmflabel{$H$}{o1}
\end{fmfgraph*}
\hspace{15mm}
\begin{fmfgraph*}(24,12)
  \fmfstraight
  \fmfleftn{i}{3} \fmfrightn{o}{3}
  \fmf{fermion}{i1,v1,i3}
  \fmf{boson,tens=3/2}{v1,v2}
  \fmf{dashes}{v2,o1} \fmflabel{$H$}{o1}
  \fmf{dashes}{v2,o2} \fmflabel{$H$}{o2}
  \fmf{boson}{v2,o3} \fmflabel{$W/Z$}{o3}
\end{fmfgraph*}
\end{fmfshrink}
}
\\[2\baselineskip]
\underline{$WW/ZZ$ double-Higgs fusion: $qq\to qqHH$}\\[1.5\baselineskip]
{\footnotesize
\unitlength1mm
\hspace{10mm}
\begin{fmfshrink}{0.7}
\begin{fmfgraph*}(24,20)
  \fmfstraight
  \fmfleftn{i}{8} \fmfrightn{o}{8}
  \fmf{fermion,tens=3/2}{i2,v1} \fmf{phantom}{v1,o2}
  \fmflabel{$q$}{i2}
  \fmf{fermion,tens=3/2}{i7,v2} \fmf{phantom}{v2,o7} 
  \fmflabel{$q$}{i7}
  \fmffreeze
  \fmf{fermion}{v1,o1} 
  \fmf{fermion}{v2,o8} 
  \fmf{boson}{v1,v3} 
  \fmf{boson}{v3,v2}
  \fmf{dashes,lab=$H$}{v3,v4}
  \fmf{dashes}{v4,o3} \fmf{dashes}{v4,o6}
  \fmflabel{$H$}{o3} \fmflabel{$H$}{o6}
  \fmffreeze
  \fmf{phantom,lab=$W/Z$,lab.s=left}{v1,x1} \fmf{phantom}{x1,v3} 
  \fmf{phantom,lab=$W/Z$,lab.s=left}{x2,v2} \fmf{phantom}{v3,x2}
  \fmfdot{v4}
\end{fmfgraph*}
\hspace{15mm}
\begin{fmfgraph*}(24,20)
  \fmfstraight
  \fmfleftn{i}{8} \fmfrightn{o}{8}
  \fmf{fermion,tens=3/2}{i2,v1} \fmf{phantom}{v1,o2}
  \fmf{fermion,tens=3/2}{i7,v2} \fmf{phantom}{v2,o7} 
  \fmffreeze
  \fmf{fermion}{v1,o1}
  \fmf{fermion}{v2,o8}
  \fmf{boson}{v1,v3} 
  \fmf{boson}{v4,v2}
  \fmf{boson,lab=$W/Z$,lab.s=left}{v3,v4}
  \fmf{dashes}{v3,o3} \fmf{dashes}{v4,o6}
  \fmflabel{$H$}{o3} \fmflabel{$H$}{o6}
\end{fmfgraph*}
\hspace{15mm}
\begin{fmfgraph*}(24,20)
  \fmfstraight
  \fmfleftn{i}{8} \fmfrightn{o}{8}
  \fmf{fermion,tens=3/2}{i2,v1} \fmf{phantom}{v1,o2}
  \fmf{fermion,tens=3/2}{i7,v2} \fmf{phantom}{v2,o7} 
  \fmffreeze
  \fmf{fermion}{v1,o1}
  \fmf{fermion}{v2,o8}
  \fmf{boson}{v1,v3} 
  \fmf{boson}{v3,v2}
  \fmf{dashes}{v3,o3} \fmf{dashes}{v3,o6}
  \fmflabel{$H$}{o3} \fmflabel{$H$}{o6}
  \fmffreeze
\end{fmfgraph*}
\end{fmfshrink}
}
\\[2\baselineskip]
\underline{$gg$ double-Higgs fusion: $gg\to HH$}\\[1.5\baselineskip]
{\footnotesize
\unitlength1mm
\hspace{10mm}
\begin{fmfshrink}{0.7}
\begin{fmfgraph*}(30,12)
  \fmfstraight
  \fmfleftn{i}{2} \fmfrightn{o}{2}
  \fmflabel{$g$}{i1}  \fmflabel{$g$}{i2}
  \fmf{gluon,tens=2/3}{i1,v1} \fmf{phantom}{v1,v2,v3,o1}
  \fmf{gluon,tens=2/3}{w1,i2} \fmf{phantom}{w1,w2,w3,o2}
  \fmffreeze
  \fmf{fermion}{w1,x2,v1}
  \fmf{dashes, lab=$H$}{x2,x3}
  \fmf{dashes}{o1,x3,o2}
  \fmffreeze
  \fmf{fermion,label=$t$,label.s=left}{v1,w1}
  \fmflabel{$H$}{o1}  \fmflabel{$H$}{o2}
  \fmfdot{x3}
\end{fmfgraph*}
\hspace{15mm}
\begin{fmfgraph*}(30,12)
  \fmfstraight
  \fmfleftn{i}{2} \fmfrightn{o}{2}
  \fmf{gluon}{i1,v1} \fmf{phantom}{v1,v3} \fmf{dashes}{v3,o1}
  \fmf{gluon}{w1,i2} \fmf{phantom}{w1,w3} \fmf{dashes}{w3,o2}
  \fmffreeze
  \fmf{fermion}{v1,w1,w3,v3,v1}
  \fmflabel{$H$}{o1}  \fmflabel{$H$}{o2}
\end{fmfgraph*}
\end{fmfshrink}
}
\end{flushleft}
\caption{\textit{%
Processes contributing to Higgs-pair production in the Standard Model
at the LHC: double Higgs-strahlung, $WW/ZZ$ fusion, and $gg$ fusion
(generic diagrams).
}}
\label{smdiag}
\end{figure}
They can be expressed in terms of the two mixing angles $\beta$ and
$\alpha$, mixing angle in the CP-even Higgs sector.  The couplings
involving two light Higgs bosons, for example, are given by the
trigonometric functions
\beq
\lambda_{hhh} &=& 3 \cos2\alpha \sin (\beta+\alpha) 
+ 3 \frac{\epsilon}{M_Z^2} \frac{\cos \alpha}{\sin\beta} \cos^2\alpha  
\\
\lambda_{Hhh} &=& 2\sin2 \alpha \sin (\beta+\alpha) -\cos 2\alpha 
\cos(\beta
+ \alpha) + 3 \frac{\epsilon}{M_Z^2} \frac{\sin \alpha}{\sin\beta}
\cos^2\alpha \non
\label{coup}
\eeq
The couplings are defined in units of $\lambda_0$. They are
renormalized indirectly by the re\-nor\-ma\-li\-za\-tion of the mixing
angle $\alpha$ and directly by additive terms proportional to the
radiative correction parameter $\epsilon$ which, to leading order, is
given by $\epsilon = 3 G_F M_t^4/(\sqrt{2}\pi^2\sin^2\beta)\cdot
\ln(M^2_{\tilde{t}}/M_t^2)$ \cite{okada}.  In the subsequent numerical
analysis the complete one-loop and the leading two-loop corrections to
the Higgs masses and couplings \cite{carena} are included . The size
of the couplings has been exemplified for a set of parameters in
Ref.\cite{hhhlc}. \s

The trilinear MSSM Higgs couplings are involved in a large number of
multi-Higgs processes at the LHC \citerange{conf,plehn}:
\beq
\begin{array}{l@{:\quad}l@{\;\to\;}l l l l}
\mbox{double Higgs-strahlung} & $\qq$  & W/Z + H_i H_j & \mathrm{and} 
& W/Z + AA & [H_{i,j}=h,H] \non\\[0.2cm]
\mbox{triple Higgs production} & $\qq$ & AH_i H_j & \mathrm{and}
& AAA & \non \\[0.2cm]
WW/ZZ\ \mbox{double-Higgs fusion} & qq & qq+ H_i H_j & \mathrm{and}
& qq+ AA & \non \\[0.2cm]
gg\ \mbox{fusion} & gg & H_i H_j,\quad H_iA & \mathrm{and}
& AA & \non
\end{array} 
\eeq
In this analysis we will restrict ourselves to a specific class of
final states while a comprehensive description of all processes will
be deferred to a subsequent report \cite{muehl}; we will study final
states involving two light Higgs bosons $h$:
\beq
pp &\to& gg \to hh \\
pp &\to& Z/W + hh \qquad {\rm and} \quad A+hh \non
\eeq
They are generated either in the continuum or, for moderate values of
$\tan\beta$, in cascade decays, too \cite{djoukal}:
\beq
H \to hh, \quad A \to Zh \quad {\rm and} \quad H^{\pm} \to W^{\pm} 
h
\eeq
A set of typical diagrams is shown in Fig.\ref{fig:mssm}. We will also
present selected results on cascade decays involving heavy Higgs
bosons $H$ in the final state; they can be generated in the chains
\beq
q\bar{q} &\to& Z^* \to AH \to ZHh \\
q\bar{q} &\to& W^* \to H^{\pm}H \to WHh \non
\eeq
These chains give rise to large rates, yet they do not involve
trilinear Higgs couplings but only gauge couplings. The corresponding
diagrams are shown in Fig.\ref{nocoup}.\s
\begin{figure}
\begin{flushleft}
\underline{double Higgs-strahlung: $q\bar q\to Zhh/Whh$}\\[1.5\baselineskip]
{\footnotesize
\unitlength1mm
\hspace{5mm}
\begin{fmfshrink}{0.7}
\begin{fmfgraph*}(24,12)
  \fmfstraight
  \fmfleftn{i}{3} \fmfrightn{o}{3}
  \fmf{fermion}{i1,v1,i3}
  \fmflabel{$q$}{i1} \fmflabel{$\bar q$}{i3}
  \fmf{boson,tens=3/2,label=$W/Z$, label.s=left}{v1,v2}
  \fmf{boson}{v2,o3} \fmflabel{$W/Z$}{o3}
  \fmf{phantom}{v2,o1}
  \fmffreeze
  \fmf{dashes,lab=$h,,H$,lab.s=right}{v2,v3} \fmf{dashes}{v3,o1}
  \fmffreeze
  \fmf{dashes}{v3,o2} 
  \fmflabel{$h$}{o2} \fmflabel{$h$}{o1}
  \fmfdot{v3}
\end{fmfgraph*}
\hspace{15mm}
\begin{fmfgraph*}(24,12)
  \fmfstraight
  \fmfleftn{i}{3} \fmfrightn{o}{3}
  \fmf{fermion}{i1,v1,i3}
  \fmf{boson,tens=3/2}{v1,v2}
  \fmf{dashes}{v2,o1}
  \fmf{phantom}{v2,o3}
  \fmffreeze
  \fmf{dashes,lab=$H^\pm/A$,lab.s=left}{v2,v3} 
  \fmf{boson}{v3,o3} \fmflabel{$W/Z$}{o3}
  \fmffreeze
  \fmf{dashes}{v3,o2} 
  \fmflabel{$h$}{o2} \fmflabel{$h$}{o1}
\end{fmfgraph*}
\hspace{15mm}
\begin{fmfgraph*}(24,12)
  \fmfstraight
  \fmfleftn{i}{3} \fmfrightn{o}{3}
  \fmf{fermion}{i1,v1,i3}
  \fmf{boson,tens=3/2}{v1,v2}
  \fmf{dashes}{v2,o1}
  \fmf{phantom}{v2,o3}
  \fmffreeze
  \fmf{boson}{v2,v3,o3} \fmflabel{$W/Z$}{o3}
  \fmffreeze
  \fmf{dashes}{v3,o2} 
  \fmflabel{$h$}{o2} \fmflabel{$h$}{o1}
\end{fmfgraph*}
\hspace{15mm}
\begin{fmfgraph*}(24,12)
  \fmfstraight
  \fmfleftn{i}{3} \fmfrightn{o}{3}
  \fmf{fermion}{i1,v1,i3}
  \fmf{boson,tens=3/2}{v1,v2}
  \fmf{dashes}{v2,o1} \fmflabel{$h$}{o1}
  \fmf{dashes}{v2,o2} \fmflabel{$h$}{o2}
  \fmf{boson}{v2,o3} \fmflabel{$W/Z$}{o3}
\end{fmfgraph*}
\end{fmfshrink}
}
\\[2\baselineskip]
\underline{triple Higgs production: $q\bar q\to Ahh$}
\\[1.5\baselineskip]
{\footnotesize
\unitlength1mm
\hspace{5mm}
\begin{fmfshrink}{0.7}
\begin{fmfgraph*}(24,12)
  \fmfstraight
  \fmfleftn{i}{3} \fmfrightn{o}{3}
  \fmf{fermion}{i1,v1,i3}
  \fmflabel{$q$}{i1} \fmflabel{$\bar q$}{i3}
  \fmf{boson,tens=3/2,label=$Z$, label.s=left}{v1,v2}
  \fmf{dashes}{v2,o3} \fmflabel{$A$}{o3}
  \fmf{phantom}{v2,o1}
  \fmffreeze
  \fmf{dashes,lab=$H,,h$,lab.s=right}{v2,v3} \fmf{dashes}{v3,o1}
  \fmffreeze
  \fmf{dashes}{v3,o2} 
  \fmflabel{$h$}{o2} \fmflabel{$h$}{o1}
  \fmfdot{v3}
\end{fmfgraph*}
\hspace{15mm}
\begin{fmfgraph*}(24,12)
  \fmfstraight
  \fmfleftn{i}{3} \fmfrightn{o}{3}
  \fmf{fermion}{i1,v1,i3}
  \fmf{boson,tens=3/2}{v1,v2}
  \fmf{dashes}{v2,o1} \fmflabel{$h$}{o1}
  \fmf{phantom}{v2,o3}
  \fmffreeze
  \fmf{dashes,lab=$A$,lab.s=left}{v2,v3} 
  \fmf{dashes}{v3,o3}
  \fmffreeze
  \fmf{dashes}{v3,o2} 
  \fmflabel{$h$}{o2} \fmflabel{$A$}{o3}
  \fmfdot{v3}
\end{fmfgraph*}
\hspace{15mm}
\begin{fmfgraph*}(24,12)
  \fmfstraight
  \fmfleftn{i}{3} \fmfrightn{o}{3}
  \fmf{fermion}{i1,v1,i3}
  \fmf{boson,tens=3/2}{v1,v2}
  \fmf{dashes}{v2,o1}
  \fmf{phantom}{v2,o3}
  \fmffreeze
  \fmf{boson}{v2,v3} 
  \fmf{dashes}{v3,o3} \fmflabel{$A$}{o3}
  \fmffreeze
  \fmf{dashes}{v3,o2} 
  \fmflabel{$h$}{o2} \fmflabel{$h$}{o1}
\end{fmfgraph*}
\end{fmfshrink}
}
\\[2\baselineskip]
\underline{$WW/ZZ$ double-Higgs fusion: $qq\to qqhh$}
\\[1.5\baselineskip]
{\footnotesize
\unitlength1mm
\hspace{5mm}
\begin{fmfshrink}{0.7}
\begin{fmfgraph*}(24,20)
  \fmfstraight
  \fmfleftn{i}{8} \fmfrightn{o}{8}
  \fmf{fermion,tens=3/2}{i2,v1} \fmf{phantom}{v1,o2}
  \fmf{fermion,tens=3/2}{i7,v2} \fmf{phantom}{v2,o7} 
  \fmflabel{$q$}{i2} \fmflabel{$q$}{i7}
  \fmffreeze
  \fmf{fermion}{v1,o1}
  \fmf{fermion}{v2,o8}
  \fmf{boson}{v1,v3} 
  \fmf{boson}{v3,v2}
  \fmf{dashes,lab=$H,,h$}{v3,v4}
  \fmf{dashes}{v4,o3} \fmf{dashes}{v4,o6}
  \fmflabel{$h$}{o3} \fmflabel{$h$}{o6}
  \fmffreeze
  \fmf{phantom,lab=$W/Z$,lab.s=left}{v1,x1} \fmf{phantom}{x1,v3} 
  \fmf{phantom,lab=$W/Z$,lab.s=left}{x2,v2} \fmf{phantom}{v3,x2}
  \fmfdot{v4}
\end{fmfgraph*}
\hspace{15mm}
\begin{fmfgraph*}(24,20)
  \fmfstraight
  \fmfleftn{i}{8} \fmfrightn{o}{8}
  \fmf{fermion,tens=3/2}{i2,v1} \fmf{phantom}{v1,o2}
  \fmf{fermion,tens=3/2}{i7,v2} \fmf{phantom}{v2,o7}
  \fmffreeze
  \fmf{fermion}{v1,o1}
  \fmf{fermion}{v2,o8}
  \fmf{boson}{v1,v3} 
  \fmf{boson}{v4,v2}
  \fmf{boson,lab=$W/Z$,lab.s=left}{v3,v4}
  \fmf{dashes}{v3,o3} \fmf{dashes}{v4,o6}
  \fmflabel{$h$}{o3} \fmflabel{$h$}{o6}
\end{fmfgraph*}
\hspace{15mm}
\begin{fmfgraph*}(24,20)
  \fmfstraight
  \fmfleftn{i}{8} \fmfrightn{o}{8}
  \fmf{fermion,tens=3/2}{i2,v1} \fmf{phantom}{v1,o2}
  \fmf{fermion,tens=3/2}{i7,v2} \fmf{phantom}{v2,o7} 
  \fmffreeze
  \fmf{fermion}{v1,o1}
  \fmf{fermion}{v2,o8}
  \fmf{boson}{v1,v3} 
  \fmf{boson}{v4,v2}
  \fmf{dashes,lab=$H^\pm/A$,lab.s=left}{v3,v4}
  \fmf{dashes}{v3,o3} \fmf{dashes}{v4,o6}
  \fmflabel{$h$}{o3} \fmflabel{$h$}{o6}
\end{fmfgraph*}
\hspace{15mm}
\begin{fmfgraph*}(24,20)
  \fmfstraight
  \fmfleftn{i}{8} \fmfrightn{o}{8}
  \fmf{fermion,tens=3/2}{i2,v1} \fmf{phantom}{v1,o2}
  \fmf{fermion,tens=3/2}{i7,v2} \fmf{phantom}{v2,o7}
  \fmffreeze
  \fmf{fermion}{v1,o1}
  \fmf{fermion}{v2,o8}
  \fmf{boson}{v1,v3} 
  \fmf{boson}{v3,v2}
  \fmf{dashes}{v3,o3} \fmf{dashes}{v3,o6}
  \fmflabel{$h$}{o3} \fmflabel{$h$}{o6}
\end{fmfgraph*}
\end{fmfshrink}
}
\\[2\baselineskip]
\underline{$gg$ double-Higgs fusion: $gg\to hh$}\\[1.5\baselineskip]
{\footnotesize
\unitlength1mm
\hspace{5mm}
\begin{fmfshrink}{0.7}
\begin{fmfgraph*}(30,12)
  \fmfstraight
  \fmfleftn{i}{2} \fmfrightn{o}{2}
  \fmflabel{$g$}{i1}  \fmflabel{$g$}{i2}
  \fmf{gluon,tens=2/3}{i1,v1} \fmf{phantom}{v1,v2,v3,o1}
  \fmf{gluon,tens=2/3}{w1,i2} \fmf{phantom}{w1,w2,w3,o2}
  \fmffreeze
  \fmf{fermion}{w1,x2,v1}
  \fmf{dashes, lab=$h,,H$}{x2,x3}
  \fmf{dashes}{o1,x3,o2}
  \fmffreeze
  \fmf{fermion,label=$t,,b$,label.s=left}{v1,w1}
  \fmflabel{$h$}{o1}  \fmflabel{$h$}{o2}
  \fmfdot{x3}
\end{fmfgraph*}
\hspace{15mm}
\begin{fmfgraph*}(30,12)
  \fmfstraight
  \fmfleftn{i}{2} \fmfrightn{o}{2}
  \fmf{gluon}{i1,v1} \fmf{phantom}{v1,v3} \fmf{dashes}{v3,o1}
  \fmf{gluon}{w1,i2} \fmf{phantom}{w1,w3} \fmf{dashes}{w3,o2}
  \fmffreeze
  \fmf{fermion}{v1,w1,w3,v3,v1}
  \fmflabel{$h$}{o1}  \fmflabel{$h$}{o2}
\end{fmfgraph*}
\end{fmfshrink}
}
\end{flushleft}
\caption{\textit{%
Processes contributing to double and triple Higgs production involving
trilinear couplings in the MSSM.
}}
\label{fig:mssm}
\end{figure}
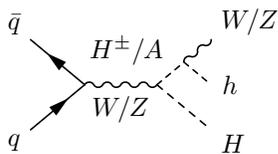
\begin{figure}
\begin{flushleft}
\underline{cascade decay: $q\bar q\to AH/HH^\pm \to ZHh/WHh$}\\[1.5\baselineskip]
{\footnotesize
\unitlength1mm
\hspace{5mm}
\begin{fmfshrink}{0.7}
\begin{fmfgraph*}(24,12)
  \fmfstraight
  \fmfleftn{i}{3} \fmfrightn{o}{3}
  \fmf{fermion}{i1,v1,i3}
  \fmflabel{$q$}{i1} \fmflabel{$\bar q$}{i3}
  \fmf{boson,tens=3/2,lab=$W/Z$,lab.s=right}{v1,v2}
  \fmf{dashes}{v2,o1}
  \fmf{phantom}{v2,o3}
  \fmffreeze
  \fmf{dashes,lab=$H^\pm/A$,lab.s=left}{v2,v3} 
  \fmf{boson}{v3,o3} \fmflabel{$W/Z$}{o3}
  \fmffreeze
  \fmf{dashes}{v3,o2} 
  \fmflabel{$h$}{o2} \fmflabel{$H$}{o1}
\end{fmfgraph*}
\end{fmfshrink}
}
\end{flushleft}
\caption{\textit{%
Processes which contribute to double light plus heavy Higgs production 
in the MSSM but which do not involve trilinear couplings.
}}
\label{nocoup}
\end{figure}

The cross sections for continuum production are generally small and it
will be difficult to discriminate the signal from the background, after
the decay of the Higgs particles into pairs of $b$ quarks, for
instance. Cascade decays, on the other hand, have been proposed to
search for these particles at the LHC \cite{richter}. \s

The present paper has got a limited goal. We have built up the general
theoretical formalism for multiple Higgs production at the LHC in the
Standard Model and the MSSM, and we discuss a few examples in detail,
see also Ref.\cite{muehl}.  Just setting the base for these processes
we do not intend to consider background reactions in a systematic way;
such simulations can only be carried out by taking proper account of 
detector properties, what is beyond the scope of this paper.

\subsection*{2. Higgs Pairs in the Standard Model}

The cross sections for double Higgs-strahlung off $W/Z$ bosons and for
vector-boson fusion can be evaluated, {\it mutatis mutandis}, at the
quark level for the LHC in the same way as for $e^+e^-$ collisions,
cf.~Ref.\cite{hhhlc}; just the couplings have to be adjusted
properly.  The proton cross sections are derived by folding the parton
cross sections $\hat{\sigma}(qq'\to HH;\hat{s})$ of the quark
subprocesses with the appropriate luminosities $d{\cal L}^{qq'}/d\tau$:
\beq
\sigma (pp\to HH) = \int_{4M_H^2/s}^1 d\tau 
\frac{d{\cal L}^{qq'}}{d\tau} 
\hat{\sigma} (qq' \to HH; \hat{s} = \tau s)
\eeq
where
\beq
\frac{d{\cal L}^{qq'}}{d\tau} = \int_\tau^1 \frac{dx}{x}
q(x;Q^2) q'(\tau/x;Q^2)
\eeq
with $q$ and $q'$ denoting the parton densities in the proton 
\cite{pdflib}, taken at a typical scale $Q\sim M_H$.\s

\begin{figure}
\begin{center}
\epsfig{figure=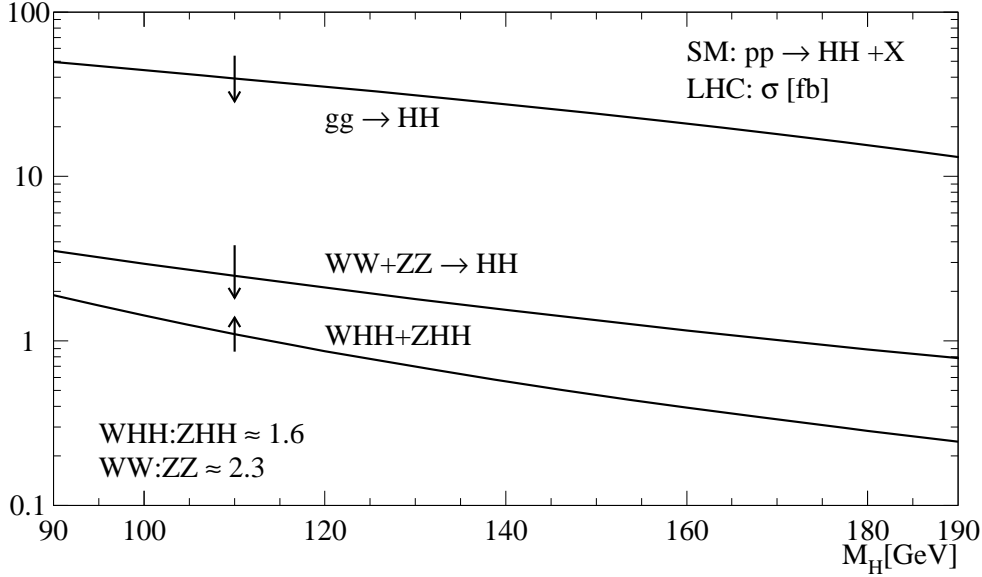,width=13cm}
\caption{The cross sections for gluon fusion, $WW/ZZ$ fusion and 
double Higgs-strahlung $WHH$, $ZHH$ in the SM. The vertical arrows 
correspond to a variation of the trilinear Higgs coupling from $1/2$ 
to $3/2$ of the SM value.}
\label{fig:SM}
\end{center}
\end{figure}
The large number of gluons in high-energy proton beams provides an
additional mechanism for the production of Higgs pairs: gluon fusion
$gg \to HH$ \cite{glover}. The proton cross section is derived by
folding the parton cross section $\hat{\sigma}(gg\to HH)$ with the
gluon luminosity. The coupling between gluons and SM Higgs bosons is
mediated by heavy top-quark loops. As expected from single Higgs
production \cite{mspira}, QCD radiative corrections are particularly
important for this channel. They have been determined in the
low-energy limit of small Higgs masses $M_H^2 \ll 4M_t^2$, leading to
a $K$ factor $K\approx 1.9$ \cite{dawson}. A $K$ factor of similar
size is generally expected for Higgs masses beyond the top-quark
threshold.\s

The cross sections are shown in Fig.\ref{fig:SM} for the intermediate
Higgs mass range discussed above. Gluon fusion dominates over the
other mechanisms. The $WW/ZZ$ fusion mechanisms are the next important
channels. In addition to the four $b$ jets, the four $W W^{(*)} W
W^{(*)}$ bosons or the mixed $bbWW^{(*)}$ pairs generated in the
decays of the two Higgs bosons, the light-quark jets associated with
the equivalent $W/Z$ bosons in the fragmentation, $q\to W/Z +q$ can be
exploited to tag fusion events; these jets are emitted at average
transverse momenta $p_{T} \sim 1/2 M_{W/Z}$. $WW$ fusion dominates
over $ZZ$ fusion at a ratio $WW$:$ZZ\approx 2.3$. The cross sections for
double Higgs-strahlung are relatively small. This follows from the
scaling behavior of the cross sections which drop $\sim 1/\hat{s}$.
The cross sections for Higgs-strahlung off $W$ and $Z$ bosons are
combined in Fig.\ref{fig:SM}; their relative size is close to
$W/Z\approx 1.6$.  The vertical arrows indicate the sensitivity of the
cross sections to the size of the trilinear Higgs coupling; they
correspond to a modification of the trilinear SM coupling
$\lambda_{HHH}$ by the {\it ad hoc} rescaling coefficient
$\kappa=1/2\to3/2$.

\subsection*{3. Higgs Pairs in Supersymmetric Theories}

With appropriate modifications, the pattern of MSSM Higgs
pair-production at the LHC is similar to the characteristics in
$e^+e^-$ collisions \cite{hhhlc,djouadi}. An important exception
however is the additional gluon-fusion channel \cite{plehn}. \s

\begin{figure}
\begin{center}
\epsfig{figure=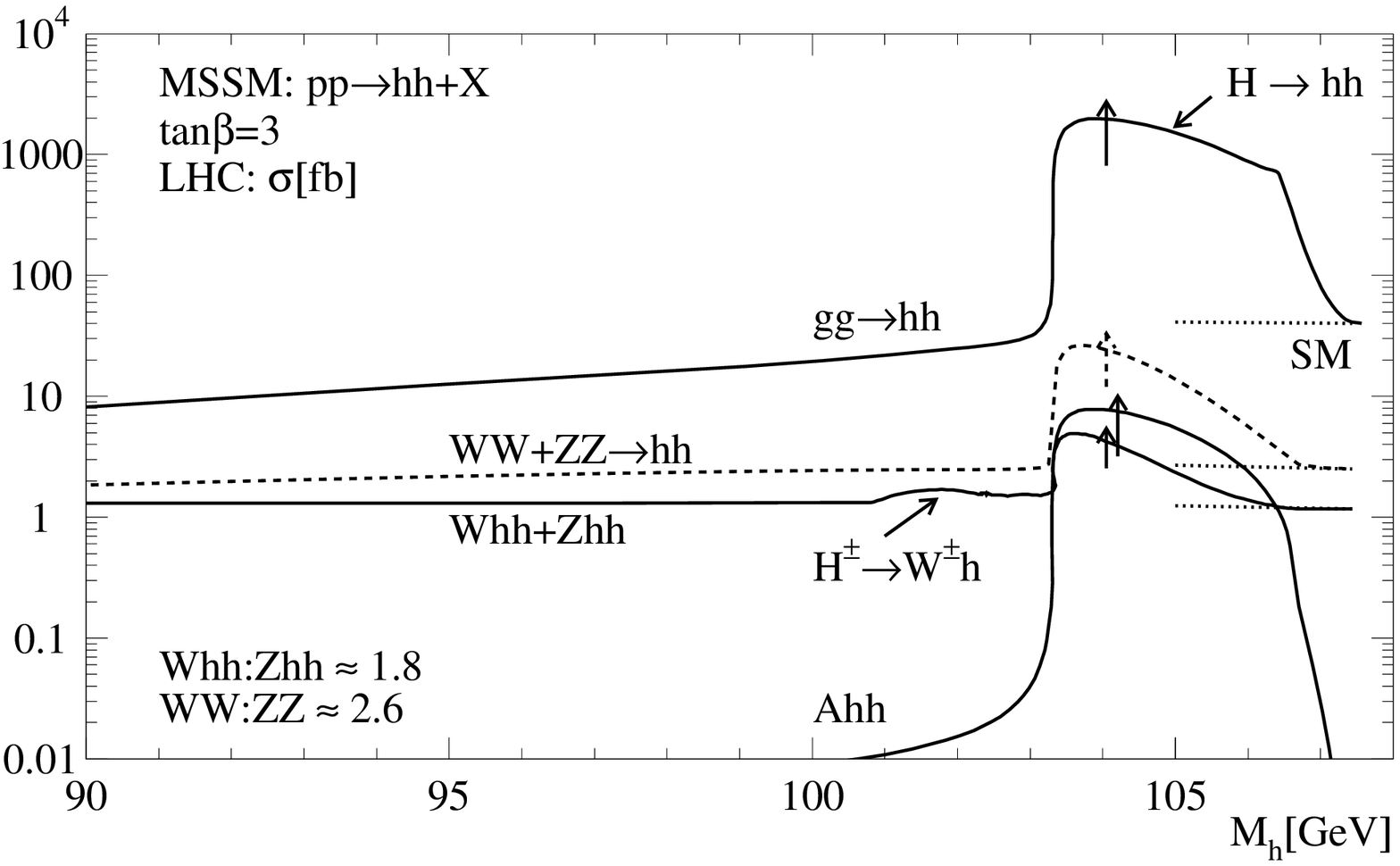,width=13cm}
\end{center}
Figure 5a:{\it Total cross sections for MSSM hh production via double 
Higgs-strahlung $Whh$ and $Zhh$, $WW/ZZ$ fusion and gluon fusion at the 
LHC for $\tan\beta=3$, including mixing effects ($A=1$~TeV, 
$\mu=-1$~TeV).}
\end{figure}
\nn 
{\bf 1.} We will focus on the production of pairs of light Higgs
bosons: $pp \to hh$. For moderate values of $\tan\beta$, the $hh$
production channels follow the pattern of the Standard Model, with
gluon fusion being dominant, Fig.5a. However, within the cascade-decay
regions of the heavy Higgs bosons $H$, $H^{\pm}$ the cross sections
rise dramatically. These domains are marked in the figures explicitly
by arrows. Large contributions to the cross sections are generated by
heavy Higgs formation $gg/VV \to H \to hh$ in the fusion channels, and
$H^{\pm} \to W^{\pm} h$ decay in Higgs-strahlung $W^{\pm*} \to H^{\pm}
h\to W^{\pm} hh$. As expected \cite{plehn}, the gluon-fusion $hh$
cross section becomes very large in the $H$ decay region, giving rise
to a sample of about a million $hh$ events. This process therefore
provides an important channel for searching for MSSM Higgs bosons at
the LHC \cite{richter}.  The sensitivity of the cross sections with
regard to a variation of $\lambda_{hhh}$ by the rescaling factor
$\kappa = 1/2$ to $3/2$ is close to ten per-cent in the continuum
while the sensitivity of $H$ cascade decays to a variation of
$\lambda_{Hhh}$ is indicated by arrows. \s

\begin{figure}
\begin{center}
\epsfig{figure=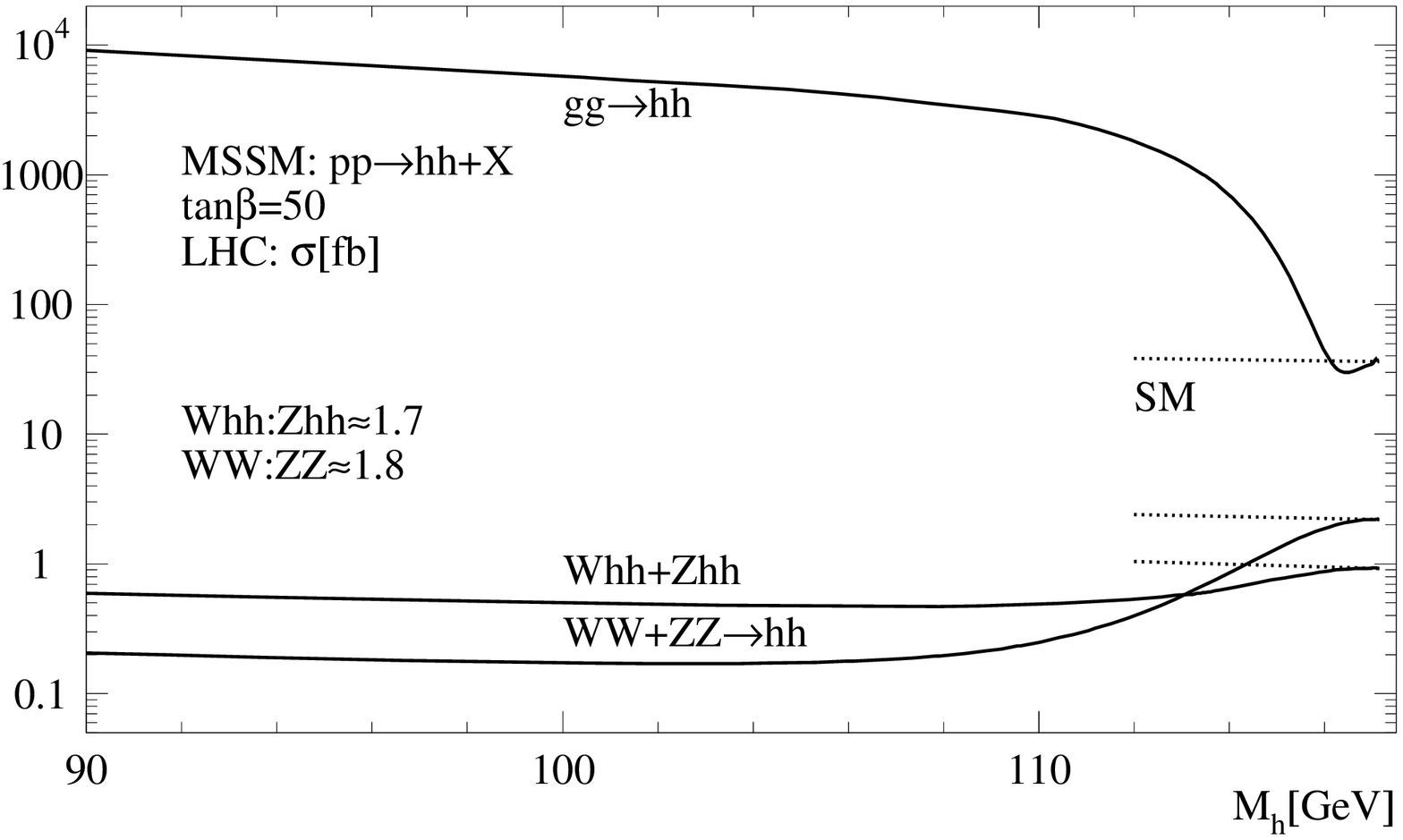,width=13cm}
\end{center}
Figure 5b:{\it Total cross sections for MSSM hh production via double 
Higgs-strahlung $Whh$, $Zhh$, $WW/ZZ$ fusion and gluon fusion at the 
LHC for $\tan\beta=50$, including mixing effects ($A=1$~TeV, 
$\mu=1$~TeV).}
\end{figure}
\setcounter{figure}{5}
For large $\tan\beta$, a huge ensemble of $hh$ continuum events is
generated by gluon fusion, Fig.5b. The enhancement is due to the
large $hbb$ Yukawa coupling, $\sim m_b \tan\beta$, in the $b$-quark
loops connecting the gluons with the Higgs bosons. Since the box
diagrams are enhanced quadratically compared to the triangle diagrams,
the sensitivity to the trilinear coupling is small.\footnote{The
  multi-$b$ final states $pp \to hh \to (b\bar{b})(b\bar{b})$ with two
  resonance structures and large transverse momenta provide an
  outstanding signature which may be exploited to search for $h$
  Higgs bosons in the range of large $\tan \beta$ (and moderate $M_A$)
  not covered hitherto at LHC in standard channels.} The continuum
cross sections of the $VV$ fusion and Higgs-strahlung channels are
suppressed with respect to the Standard Model until the decoupling
limit is reached.  For large $\tan\beta$ cascade decays do not play a
role in $hh$ pair production; the kinematical decay thresholds are
reached only for masses for which the decoupling limit is being
approached. \pskip

\begin{figure}
\begin{center}
\epsfig{figure=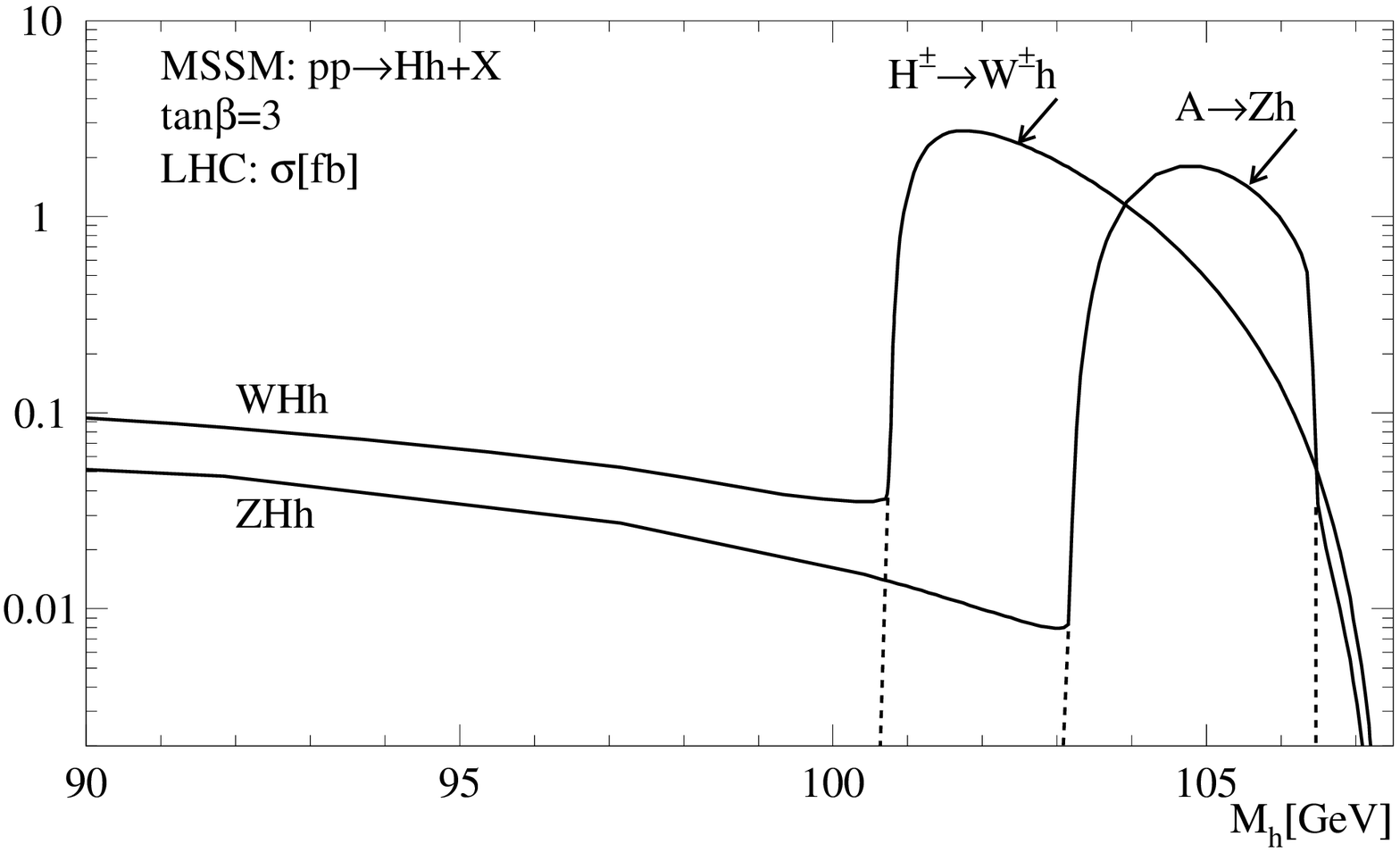,width=13cm}
\caption{\it Total cross sections for MSSM $Hh$ production in the 
processes $WHh$ and $ZHh$ for $\tan\beta=3$, including mixing effects 
($A=1$~TeV, $\mu=-1$~TeV).}
\label{heavy}
\end{center}
\end{figure}
\nn
{\bf 2.} Two processes involve Higgs-pair final states including a
light plus a heavy Higgs boson. However for cross sections in excess
of 1~fb, Fig.\ref{heavy}, the final states are generated in cascade
decays by gauge interactions:
\beq
\begin{array}{l l l l l}
pp & \hspace{-0.1cm} \to & AH &  
\hspace{-0.3cm} \to &\hspace{-0.1cm}  ZHh \\
\\[-0.8cm]
& \hspace{-0.1cm} \scriptstyle{Z}& & &  \\[0.1cm]
pp & \hspace{-0.1cm} \to & H^{\pm} H & 
\hspace{-0.3cm} \to & \hspace{-0.1cm} W^{\pm} Hh
\\ \\[-0.8cm]
& \hspace{-0.1cm} \scriptstyle{W}& & &
\end{array} 
\eeq
These processes are therefore not suitable for measuring trilinear 
Higgs couplings.

\subsection*{4. Conclusions}

In the present paper we have analyzed the production of neutral
Higgs-bosons pairs in various channels at the LHC which can eventually
be used to measure fundamental trilinear Higgs self-couplings. In a
first step we have compared the production cross sections in the
Standard Model, assuming a Higgs-boson mass in the intermediate range.
Moreover, we have calculated the cross sections for pairs of light
Higgs bosons in the Minimal Supersymmetric extension of the Standard
Model. Earlier results haven been combined with new calculations in
these analyses.\s

The continuum cross sections are generally small in the SM and MSSM,
yet not for gluon fusion. The trilinear SM Higgs coupling and the
trilinear coupling $\lambda_{hhh}$ in the MSSM may thus be accessible
experimentally, provided the backgrounds can be rejected sufficiently
well. If Higgs cascade decays $H\to hh$ occur in the MSSM, they can be
exploited to measure one of the couplings between light and heavy
CP-even neutral Higgs bosons, $\lambda_{Hhh}$. \pskip

\subsubsection*{Acknowledgements}

We are grateful to M.~Spira for discussions and for providing us with a
source code for gluon fusion of Higgs pairs.

\end{fmffile}
\newpage


\end{document}